\documentclass[%
 reprint,
nofootinbib,
 amsmath,amssymb,
 aps,
]{revtex4-1}

\usepackage{graphicx}
\usepackage{dcolumn}
\usepackage{bm}
\usepackage{amssymb}
\usepackage{amsmath}
\usepackage{amstext}
\usepackage{comment}
\usepackage{hyperref}

\newcommand{\ba}{\begin{eqnarray}}
\newcommand{\ea}{\end{eqnarray}}
\newcommand{\be}{\begin{equation}}
\newcommand{\ee}{\end{equation}}
\newcommand{\bi}{\begin{itemize}}
\newcommand{\ei}{\end{itemize}}
\newcommand{\al}{\alpha}

\newcommand{\ga}{\gamma}

\newcommand{\da}{\delta}
\newcommand{\la}{\lambda}

\newcommand{\sa}{\sigma}

\newcommand{\cP}{{\cal P}}

\newcommand{\p}{\partial}

\newcommand{\ra}{\rightarrow}

\newcommand{\LF}{\left(}
\newcommand{\RF}{\right)}
\newcommand{\LT}{\left[}
\newcommand{\RT}{\right]}
\newcommand{\Ld}{\left.}
\newcommand{\Rd}{\right.}

\newcommand{\non}{\nonumber\\}

\begin{document}


\title{Projecting the graviton to probe higher or lower dimensions   }

\author{Anupam Mazumdar~$^{1}$}
\author{Spyridon Talaganis~$^{1}$}
\affiliation{
$^{1}$~Consortium for Fundamental Physics, Lancaster University, Lancaster, LA1 4YB, UK}

\date{\today}

\begin{abstract}
In this paper, we will discuss how in Einstein's theory a graviton can be enforced to probe fewer space-time dimensions in the deep ultraviolet (UV) as compared to far infrared (IR), and vice-versa. In particular, from a $D$ dimensional gravitational action in the IR
we can project the $D$ dimensional graviton to probe only $2\leq N\leq D-1$ dimensions in the deep UV.
Such projections of a graviton propagator can be thought of as an alternative to compactification. We will briefly explain the physical interpretation and consequences of such a dimensional transmutation.
\end{abstract}

\maketitle

\section{\label{sec:intro}Introduction}

Einstein's general theory of relativity has very interesting predictions in the infrared (IR) in $4$ dimensions, which has matched very well with the current observations~\cite{will}. In $4$ dimensions, the linearised gravity can be quantised~\cite{Gupta}, and can be described by the $2$ massless degrees of freedom. In the ultraviolet (UV), at short distances and at small time scales, the Einstein's theory of gravity has challenges - the classical solutions exhibit blackhole and cosmological type singularities, and furthermore at a quantum level the theory is not UV finite. It is believed that perhaps super string theory would address some quantum aspects of gravity, but in higher dimensions~\cite{Polchinski:1998rr}, and there are other compelling approaches such as loop quantum gravity~\cite{Abhay},  dynamical triangulation~\cite{Ambjorn}, asymptotic safety~\cite{Weinberg}, or 
non-local singularity free gravity in $4$ dimensions~\cite{Tomboulis,Biswas}.

Indeed, some of the UV aspects of gravity can be ameliorated if we had to lower the space-time dimensions, such as in $3$ dimensions
where quantum gravity is exactly solvable~\cite{Witten}, and in $2$ dimensions where the theory possesses no dynamical scale. The graviton
in $2$ space-time dimensions does not have any propagating degree of freedom, as a result the theory becomes a true quantum problem of nature. It is then indeed wishful to think how our Universe would emerge from $2$ space-time dimensions to that  of $4$ dimensions~\cite{Brandenberger:1988aj}.

Addressing some of these conceptual issues of gravity also rely on dimensional reduction of space-time, which  is a phenomenon where a physical theory behaves as a $D$-dimensional in a specific energy regime, whereas it behaves as $N$-dimensional in another energy regime. For instance, in Kaluza-Klein (KK) theory~\cite{Kaluza,Klein} or in string theory, dimensional reduction denotes a situation where a theory is higher-dimensional in the UV, but four-dimensional in the IR, below a certain scale, sometimes known as the compactification scale. There are also alternatives to compactification, such as in Randall-Sundrum scenarios~\cite{Randall:1999vf}, where a higher-dimensional graviton wave-function can be projected on to a brane embedded in a higher dimensional theory.

The aim of this paper is very simple; we wish to realise a graviton propagator which can undergo dimensional transmutation from $D$ space-time dimensions in the far IR to say $2 \leq N \leq D-1$ space-time dimensions in the deep UV. We will measure the UV and IR in terms of $D$ dimensional theory, in particular the $D$ dimensional momentum will determine UV and IR regimes.

Our underlying gravitational action remains the Einstein-Hilbert action, but we wish to control the dynamical degrees of freedom of a graviton in such a way that, in the UV, some or all the dynamical degrees of freedom freeze for instance. 

Typically, in any $D$ dimensional Einstein-Hilbert action, there are total $D(D-1)/2$ degrees of 
freedom~\cite{Nieuwenhuizen,Conroy:2015nva}, the additional $D$ degrees of freedom in a symmetric tensor field are gauge degrees of freedom. ${\cal P}^{(2)}$ and ${\cal P}^{(1)}$ represent the sum of the transverse and traceless spin-$2$ and spin-$1$ degrees, accounting for $(D^{2}-D-4)/2$ degrees of freedom, while ${\cal P}_s^{(0)}$ and ${\cal P}_w^{(0)}$ represent the spin-$0$ scalar multiplets. In addition to the above four spin projector operators, we have 
${\cal P}^{(0)}_{sw}$ and ${\cal P}^{(0)}_{ws}$, which mix. In particular, we will illustrate how  ${\cal P}^{(2)}$ and ${\cal P}_s^{(0)}$ 
components can be used to transmute the graviton propagation from $D$ to $N$ dimensions, and vice-versa.

Emulating similar mathematical construction would also {\it mimic} traditional KK-reduction, {\it i.e.}, from $D$ dimensions in deep UV to 
lower $2\leq N \leq D-1$ dimensions in the IR. We will also highlight some of the challenges in restricting the dynamical degrees of freedom of a graviton, and end our paper on a speculative note, which would require further understanding and development of certain concepts.

\section{Propagator for $D$-dimensional Einstein-Hilbert action}

As a warm-up exercise we will first highlight how to obtain a graviton propagator in $D$ dimensions for the 
Einstein-Hilbert action~\footnote{In this paper, we are using the ``mostly plus'' metric signature when we are working in Minkowski space.}:
\be
S_{EH}= \frac{1}{16 \pi G_{D}} \int d^{D} x \, \sqrt{-g} R\,,
\ee
where $G_{D}$ is the $D$-dimensional Newton's constant. One can linearise the action around 
the Minkowski space-time, with $g_{\mu\nu}=\eta_{\mu\nu}+h_{\mu\nu}$, where $\mu,~\nu =0,\dots,D-1$, 
and keep the quadratic order terms, {\it i.e.},  ${\cal O}(h^2)$ terms.
The $\mathcal{O}(h^2)$ part of the action can be written as:
\begin{align}
S_{q} & =- \frac{1}{32 \pi G_{D}} \int d^{D} x \, \left[\frac{1}{2}h_{\mu \nu} \Box h^{\mu
\nu}-h_{\mu}^{\sigma}  \partial_{\sigma} \partial_{\nu} h^{\mu \nu}
\Rd \non
& + \Ld  h\partial_{\mu} \partial_{\nu}h^{\mu \nu} -\frac{1}{2}h \Box h \right]\,.
\end{align}
The field equations are given by:
\be \label{eq:tuku}
\Box h_{\mu \nu}-\p_{\sa}\p_{(\nu}h_{\mu)}^{\sa}+\eta_{\mu \nu}\p_{\rho}\p_{\sa}h^{\rho \sa}+\p_{\mu}\p_{\nu}h-\eta_{\mu \nu}\Box h=-\kappa \tau_{\mu \nu}\,,
\ee
for an energy-momentum tensor $\tau_{\mu\nu}$ and $\kappa =16 \pi G_{D}$.
The field equations can also be written in the form
\be
\Pi_{\mu \nu}^{-1 \la \sa}h_{\la \sa}=\kappa \tau_{\mu \nu}\,,
\ee
where $\Pi_{\mu \nu}^{-1 \la \sa}$ is the inverse propagator.

The inverse propagator $\Pi^{-1}$ can be expressed in terms of six spin projector operators, $\mathcal{P}^{i}$, {\it i.e.}, see~\cite{Nieuwenhuizen,Conroy:2015nva,Biswas:2013kla}:
\be
\Pi^{-1}=\sum_{i=1}^{6}\mathcal{C}_{i}\mathcal{P}^{i}\,,
\ee
where the coefficients $\mathcal{C}_{i}$'s are, in momentum space, scalars depending on $k^2$.
Note that the operators ${\cal P}^i$ are in fact $4$-rank tensors, ${\cal P}^{i}_{\mu\nu\rho\sigma}$, but here we 
have suppressed the index notation. Out of the six operators, four of them, ${\cal P}^{(2)},~{\cal P}^{(1)},~{\cal P}_s^{(0)},~{\cal P}_w^{(0)}$, 
form a complete set of projector operators:
\be
{\cal P}^{i}_{a} {\cal P}_{b}^{j} =\delta^{ij}\delta_{ab}{\cal P}_{a}^{i},~~~ {\cal P}^{(2)}+{\cal P}^{(1)}+{\cal P}_{s}^{(0)}+{\cal P}_{w}^{(0)} =1\,;
\ee
the spin projector operators are given by:
\ba
\mathcal{P}^{(2)}&=&\frac{1}{2}(\theta_{\mu \rho}\theta_{\nu \sigma}+\theta_{\mu
\sigma}\theta_{\nu \rho} ) - \frac{1}{D-1}\theta_{\mu \nu}\theta_{\rho
\sigma}\,, \nonumber \\
\mathcal{P}^{(1)}&=&\frac{1}{2}( \theta_{\mu \rho}\omega_{\nu \sigma}+\theta_{\mu
\sigma}\omega_{\nu \rho}+\theta_{\nu \rho}\omega_{\mu \sigma}+\theta_{\nu
\sigma}\omega_{\mu \rho} )\,,\nonumber \\
\mathcal{P}_{s}^{(0)}&=&\frac{1}{D-1}\theta_{\mu \nu} \theta_{\rho \sigma}\,,~~
\mathcal{P}_{w}^{(0)}=\omega_{\mu \nu}\omega_{\rho \sigma}\,, \nonumber \\
\mathcal{P}_{sw}^{(0)}&=&\frac{1}{\sqrt{D-1}}\theta_{\mu \nu}\omega_{\rho \sigma}\,,~
\mathcal{P}_{ws}^{(0)}=\frac{1}{\sqrt{D-1}}\omega_{\mu \nu}\theta_{\rho \sigma}\,,~
\ea
where
\be \label{theta}
\theta_{\mu \nu}=\eta_{\mu \nu}-\frac{k_{\mu}k_{\nu}}{k^2}\,,~~~~~~~~~~
\omega_{\mu \nu}=\frac{k_{\mu}k_{\nu}}{k^2}\,.
\ee
The operators $\cP^i$ are $4$-rank tensors, $\cP_{\mu \nu \rho \sa}^{i}$, but we shall suppress the index notation hereafter.
These projector operators now represent the $D(D-1)/2$ field degrees of freedom.  As stated earlier, the additional $D$ fields in a symmetric tensor field are gauge degrees of freedom. ${\cal P}^{(2)}$ and ${\cal P}^{(1)}$ represent the sum of the transverse and traceless spin-$2$ and spin-$1$ degrees, 
accounting for $(D^{2}-D-4)/2$ field degrees of freedom~\footnote{In a transverse-traceless decomposition of the metric tensor, there are $D(D-3)/2$ spin-$2$ helicity states and $D-2$ spin-$1$ helicity states; hence, the sum of the spin-$2$ and spin-$1$ degrees of freedom is $(D^{2}-D-4)/2$. When $D=2$, the Einstein-Hilbert action is purely topological and there exist no equations of motion, which is why we get $-1$ spin-$2$ polarisation for the graviton.}, while ${\cal P}_s^{(0)}$ and ${\cal P}_w^{(0)}$ represent the spin-$0$ scalar multiplets. 
In addition to the above four spin projector operators, we have ${\cal P}^{(0)}_{sw}$ and ${\cal P}^{(0)}_{ws}$,  which 
mix the two scalar multiplets. One can also check that the following relations are satisfied: $\theta_{\mu\rho}\theta_{\nu}^{\rho}=\theta_{\mu\nu}$, $\omega_{\mu\rho}\omega_{\nu}^{\rho}=\omega_{\mu\nu}$ and $\theta_{\mu\rho}\omega_{\nu}^{\rho}=0$. 

The transverse and traceless components of the graviton propagator  in $D$ dimensions can be recast in terms of the spin 
projector operators, which involves the tensor ${\cal P}^{(2)} $ and only one  of the scalar component ${\cal P}_s^{(0)}$:
\be
\Pi^{D}(k^2) \sim \frac{1}{k^2} \LF \mathcal{P}^{(2)}-\frac{1}{D-2}\mathcal{P}_{s}^{(0)} \RF\,,
\ee
where $k^{\mu}$ is the $D$-momentum vector ($k^{2}=-k_{0}^{2}+k_{1}^{2}+\dots+k_{D-1}^{2}$). 

In $D=4$, there exists a scalar component in the propagator, while  in $D=2$, there is no scalar component left to propagate, {\it i.e.}, the ${\cal P}_{s}^{(0)}$ spin projector operator completely decouples from the graviton dynamics, {\it i.e.}, $\Pi^{2}(k^2)\sim {\cal P}^{(2)}/k^2$.

\section{$D$ dimensions in the IR, $2\leq N\leq D-1$ dimensions in the UV}\label{sec:dukduk}

The question we are interested in asking is whether there is a way to continuously deform the graviton propagator from $D$ dimensions in the IR to $N$ dimensions in the UV. This could be achieved in a very simple way by modifying the spin projector operators~\footnote{We should point out that modifying the spin projector operators breaks general covariance; in the UV ($k_{D}^{2} \ra \infty$ in Euclidean space) and IR  ($k_{D}^{2} \ra 0$ in Euclidean space) limits, general covariance is restored.} $\cP^{(2)}$ corresponding to 
the spin-$2$ part, and  $\cP_{s}^{(0)}$, which corresponds to the $s$-multiplet, to 
\ba
&& \cP^{(2)'}=B(k^{\mu}){\cal P}^{(2)}\,,\non
&& \cP_{s}^{(0)'}=\frac{B(k^{\mu})}{A(k_{D}^{2})}{\cal P}_{s}^{(0)}\,,
\ea
where $\mu=0,1, 2,\cdots ,D-1$ and
\be
B(k_{D}^2)=\frac{k_{D}^{2}}{k_{N}^{2}+h(k_{D}^{2})(k_{D}^{2}-k_{N}^{2})}\,, 
\ee
while the other four spin projector operators stay the same, and $A(k_{D}^2)$ is a $D$-dimensional momentum dependent analytic function.

We have now the following relationship~\footnote{For the purposes of this paper, we use the notation $k_{D}^{2}=k^{\mu}k_{\mu}=-k_{0}^{2}+k_{1}^{2}+ \dots+k_{D-1}^{2}$ and $k_{N}^{2}=-k_{0}^{2}+k_{1}^{2}+\dots+k_{N-1}^{2}$.}:
\ba
&&\left[\frac{1}{B(k^{\mu})}\left({\cal P}^{(2)'}+A(k_{D}^2)\cP_{s}^{(0)'}\right)+  {\cal P}^{(1)}+{\cal P}_{w}^{(0)}\right]_{\mu\nu,\rho\sigma}\non
&& =\frac{1}{2}\left(\eta_{\mu\rho}\eta_{\nu\sigma}+\eta_{\mu\sigma}\eta_{\nu\rho}\right)
 \equiv I_{\mu\nu,\rho\sigma}. 
\ea
 In addition, we will also have the relationships:
 \begin{widetext}
\ba 
&& \mathcal{P}^{(2)'}\mathcal{P}^{(2)'}=B(k^{\mu})\mathcal{P}^{(2)'}\,, ~~~~~~  \mathcal{P}^{(1)}\mathcal{P}^{(1)}=\mathcal{P}^{(1)}\,,\\
&& \mathcal{P}^{(2)'}\mathcal{P}^{(1)}=\mathcal{P}^{(1)}\mathcal{P}^{(2)'}=0  \,, \\ 
&& \mathcal{P}_{s}^{(0)'}\mathcal{P}^{(i)}=\mathcal{P}^{(i)}\mathcal{P}_{s}^{(0)'}  =\mathcal{P}_{w}^{(0)}\mathcal{P}^{(i)}=\mathcal{P}^{(i)}\mathcal{P}_{w}^{(0)}  =0, \\
&& \mathcal{P}_{s}^{(0)'} \mathcal{P}_{s}^{(0)'}  =\frac{B(k^{\mu})}{A(k_{D}^2)}\mathcal{P}_{s}^{(0)'}\,,~~~~~ \mathcal{P}_{w}^{(0)}\mathcal{P}_{w}^{(0)}=\mathcal{P}_{w}^{(0)}\,, \\
&& \mathcal{P}_{s}^{(0)'}\mathcal{P}_{w}^{(0)}=\mathcal{P}_{w}^{(0)}\mathcal{P}_{s}^{(0)'}=0  \,, \\ 
&& \mathcal{P}_{ij}^{(0)} \mathcal{P}_{s}^{(0)'}  =\frac{B(k^{\mu})}{A(k_{D}^2)}\da_{js}\mathcal{P}_{ij}^{(0)}\,,~~~~~ \mathcal{P}_{ij}^{(0)} \mathcal{P}_{w}^{(0)}  =\da_{jw}\mathcal{P}_{ij}^{(0)}\,,  \\  
&& \mathcal{P}_{s}^{(0)'} \mathcal{P}_{ij}^{(0)}  =\frac{B(k^{\mu})}{A(k_{D}^2)}\da_{is}\mathcal{P}_{ij}^{(0)}\,,~~~~~\mathcal{P}_{w}^{(0)} \mathcal{P}_{ij}^{(0)}  =\da_{iw}\mathcal{P}_{ij}^{(0)}\,,\\
&& \mathcal{P}_{ij}^{(0)} \mathcal{P}_{sw}^{(0)}  =\da_{iw}\da_{js}\frac{A(k_{D}^2)}{B(k^{\mu})}\mathcal{P}_{s}^{(0)'},~~~~~\mathcal{P}_{ij}^{(0)} \mathcal{P}_{ws}^{(0)}  =\da_{is}\da_{jw}\mathcal{P}_{w}^{(0)}.~~~~~
\ea
Now, making use of the modified spin projector operators, we can write the field equations, see Eq.~\eqref{eq:tuku}, as
\be \label{eq:fe}
\sum_{i=1}^6 \mathcal{C}_{i}\mathcal{P}^{i}h=\kappa \left(\frac{1}{B(k^{\mu})}\mathcal{P}^{(2)'}+\mathcal{P}^{(1)}+\frac{A(k_{D}^2)}{B(k^{\mu})}\mathcal{P}_{s}^{(0)'}+\mathcal{P}_{w}^{(0)} \right)\tau,
\ee
\end{widetext}

Then, going to a momentum space, we have, for the field equations in Eq.~\eqref{eq:tuku},
\begin{widetext}
\ba
&&h_{\mu \nu} \Longrightarrow  \left[\frac{1}{B(k^{\mu})}\mathcal{P}^{(2)'}+\mathcal{P}^{(1)}+\frac{A(k_{D}^2)}{B(k^{\mu})} \mathcal{P}_{s}^{(0)'}+\mathcal{P}_{w}^{(0)}  \right]h, \qquad \\
&&-\partial_{\sigma}\partial_{(\nu}h_{\mu)}^{\sigma} \Longrightarrow k_{D}^2\left[\mathcal{P}^{(1)}+2\mathcal{P}_{w}^{(0)}\right]h,
\ea
\ba
&&\eta_{\mu \nu}\partial_{\rho}\partial_{\sigma}h^{\rho \sigma}+\partial_{\mu}\partial_{\nu}h
 \Longrightarrow
-k^2 \left[\vphantom{\sqrt{(D-1)g(k^2)}\left(\mathcal{Q}_{sw}^{0}+\mathcal{P}_{ws}^{0}\right)}2\mathcal{P}_{w}^{(0)}+ 
 \sqrt{D-1}\left(\mathcal{P}_{sw}^{(0)}+\mathcal{P}_{ws}^{(0)}\right)\right]h,
\ea
\ba
- \eta_{\mu \nu} h  \Longrightarrow - \left[\vphantom{\sqrt{(D-1)g(k^2)}\left(\mathcal{P}_{sw}^{0}+\mathcal{P}_{ws}^{0}\right)}\frac{A(k_{D}^2)}{B(k^{\mu})}(D-1)\mathcal{P}_{s}^{(0)'}+\mathcal{P}_{w}^{(0)} \Rd
 + \Ld \sqrt{D-1}\left(\mathcal{P}_{sw}^{(0)}+\mathcal{P}_{ws}^{(0)}\right)
\right]h. \qquad
\ea
By acting $\mathcal{P}^{(2)'}$ on Eq.~\eqref{eq:fe}, we obtain
\be
k_{D}^2\mathcal{P}^{(2)'}h=\kappa \mathcal{P} ^{(2)'} \tau \Longrightarrow \mathcal{P}^{(2)'}h=\kappa\left(\frac{\mathcal{P}^{(2)'}}{k_{D}^2}\right)\tau.
\ee
Similarly, acting $\mathcal{P}^{(1)}$ on Eq.~\eqref{eq:fe}, we find
\be
(1-1)k_{D}^2 \mathcal{P}^{(1)}h=\kappa\mathcal{P}^{(1)}\tau \Longrightarrow \mathcal{P}^{(1)}\tau=0.
\ee
There are no vector degrees of freedom which propagate in Einstein-Hilbert action. Hence, the vectorial part of the stress-energy tensor vanishes
identically. By acting $\mathcal{P}_{s}^{(0)'}$ and $\mathcal{P}_{w}^{(0)}$ on Eq.~\eqref{eq:fe}, we obtain
\ba \label{eq:ok}
 (1-(D-1))k_{D}^2\mathcal{P}_{s}^{(0)'}h+(1-1)k_{D}^{2} \frac{B(k^{\mu})\sqrt{D-1}}{A(k^2)}\mathcal{P}_{sw}^{(0)}h  = \kappa \mathcal{P}_{s}^{(0)'}\tau, \qquad \, \, \, \, \\
(1-1)k_{D}^2\sqrt{D-1}\mathcal{P}_{ws}^{(0)}h+(1-2+2-1)k_{D}^2\mathcal{P}_{w}^{(0)}h = \kappa \mathcal{P}_{w}^{(0)}\tau. \qquad \, \, \, \, \,
\ea
By applying the projector $\mathcal{P}_{w}^{(0)}$ on Eq.~\eqref{eq:ok}, the scalar multiplets decouple, yielding
\ba
 (1-(D-1))k_{D}^2\mathcal{P}_{s}^{(0)'}h= \kappa \mathcal{P}_{s}^{(0)'}\tau 
 \Longrightarrow \mathcal{P}_{s}^{(0)'}h=\kappa \frac{\mathcal{P}_{s}^{(0)'}}{(2-D)k_{D}^2}\tau,\\
(1-2+2-1)k_{D}^{2}\mathcal{P}_{w}^{(0)}h= \kappa \mathcal{P}_{w}^{(0)} \tau 
 \Longrightarrow \mathcal{P}_{w}^{(0)} h= \kappa \frac{\mathcal{P}_{w}^{(0)}}{(1-2+2-1)k_{D}^2}
\tau.
\ea
The denominator corresponding to the $\mathcal{P}_{w}^{(0)}$ spin projector vanishes so that there is no $w$-multiplet. 
Hence, the propagator retains the form
\ba \label{eq:rara}
 \Pi(k^{\mu}) \sim \frac{1}{k_{D}^{2}} \LF \mathcal{P}^{(2)'}-\frac{1}{D-2}\mathcal{P}_{s}^{(0)'} \RF 
 \sim \frac{1}{k_{N}^{2}+h(k_{D}^{2})(k_{D}^{2}-k_{N}^{2})} \LF \mathcal{P}^{(2)}-\frac{1}{(D-2)A(k_{D}^{2})}\mathcal{P}_{s}^{(0)} \RF\,.
\ea
\end{widetext}
 %
where
\be \label{eq:duk3}
A(k_{D}^{2})=\frac{D-N}{D-2}g(k_{D}^{2}) +\frac{N-2}{D-2}\,
\ee
is an entire function with no zeroes and $h(k_{D}^{2})$ is, again, an entire function with no zeroes. Let us remind ourselves that Eq.~\eqref{eq:rara} is written in Minkowski spacetime (``mostly plus'' metric signature). If we analytically continue Eq.~\eqref{eq:rara} to Euclidean space, it will have the same form ($k^{2} \ra k_{E}^{2}$ and we drop the subscript $E$ hereafter). Working now in Euclidean space for the rest of Section~\ref{sec:dukduk}, we want the following limits to hold:
\begin{eqnarray}\label{limits}
h(k_{D}^{2})\ra 1,~~{\rm for}~~ k_{D}^{2} \ra 0\,,\nonumber \\
h(k_{D}^{2}) \ra 0,~~{\rm for} ~~k_{D}^{2} \ra \infty\,. 
\end{eqnarray}

For the purpose of illustration, we may assume~\footnote{We may also assume {\it infinitely differentiable} entire functions with rapid decay for 
$g,~h$,~see~\cite{Tomboulis:2015gfa}.}:
\be\label{limits-form}
g(k_{D}^{2}) \sim h(k_{D}^{2})=e^{-k_{D}^{2}/M^{2}}\,.
\ee
In the UV, we see that $A(k_{D}^{2}) \ra \frac{N-2}{D-2}$ and, in the IR, $A(k_{D}^{2}) \ra 1$. We will discuss the importance of the new scale $M$ below.

This is in a way a filtering mechanism, which allows the graviton to probe less dimensions in the UV, beyond the scale $M$,
for $k^2_{D} \ra \infty$, without modifying the Einstein-Hilbert gravitational action. In principle, this could be viewed as a novel dynamical mechanism to effectively forbid the graviton to see all the space-time dimensions in the deep UV.

Regarding the expression $k_{N}^{2}+h(k_{D}^{2})(k^{2}-k_{N}^{2})$ appearing in Eq.~\eqref{eq:rara}, as $k_{D}^{2}\ra \infty$, the aforementioned expression tends to $k_{N}^{2}$, whereas, when $k_{D}^2\ra 0$, in the IR limit we recover all the $D$ dimensions 
in a graviton propagator. Our analysis clearly shows that the projectors ${\cal P}^{(2)}$ and ${\cal P}_{s}^{(0)}$ can be now 
deformed continuously. So far we have discussed in extreme limits, we will discuss what happens when the momentum modes $k_{D}^2 \sim M^2$, later on.

\subsection{From $4\ra 2$ (IR $\ra$ UV) dimensions}\label{sec:kukkuk}

As an illustrative example, let us examine a current scenario where the graviton is propagating in $4$ space-time dimensions (in subsection~\ref{sec:kukkuk}, we are, again, working in Euclidean space). 
However,  as we go higher in energies, {\it i.e.}, as the $4$-dimensional $k_{4}^2\ra \infty$, instead of $3$ spatial dimensions, we may enforce the 
graviton to probe only $1$ spatial dimension. 

In the case where $D=4$ and $N=2$, we obtain the physical graviton propagator in the IR while, in the UV, the $\mathcal{P}_{s}^{0}$ multiplet decouples, yielding the two-dimensional propagator. In the deep UV, the
graviton effectively probes the $2$ space-time dimensions, while, in the far IR, the graviton sees the full space-time. 

The interpretation of the new scale $M$ is very similar to the KK compactification scale and could be of the order of $M_{p}$. In fact, the dimensionless ratio 
\be
\alpha \sim \frac{M_{p}^{2}}{M^{2}}\,,
\ee
where $M^{2}_{p}= 1/(8\pi G_{4})$, will govern the dimensionless gravitational constant in $2$ space-time dimensions ($\ga_{ab}$ is the metric in $2$ dimensions and $R^{(2)}$ is the two-dimensional Ricci scalar), 
\be
S= \al \int_{\cal M} d^2\xi \, \sqrt{-\ga} \, R^{(2)}(\ga)\,,
\ee
which is the classical reparameterisation-invariant Einstein-Hilbert action in $2$ dimensions.

\section{From $D\ra N$ (UV $\ra$ IR )  }\label{sec:tuktuk}

Now, let us mimic the  traditional KK reduction, where our starting point is $D$ dimensional gravitational action which reduces to $1\leq N \leq D-1$ dimensions upon compactifying extra dimensions at a compactification scale, say $M$. However, now, we will
enforce the graviton to probe fewer dimensions in the IR as compare to the UV.

Emulating the prescription outlined above (in Section~\ref{sec:tuktuk}, we are working in Euclidean space), if we again assume 
$A(k_{D}^{2})$ is an entire function with no zeroes given by Eq.~\eqref{eq:duk3} and $h(k_{D}^{2})$ is, again, an entire function with no zeroes.

However, now we may assume instead of Eq.~(\ref{limits}):
\begin{eqnarray}
h(k_{D}^{2})\ra 0~~{\rm for}~~ k_{D}^{2} \ra 0\,, \nonumber \\
h(k_{D}^{2}) \ra 1~~{\rm for}~~k_{D}^{2} \ra \infty\,. 
\end{eqnarray}
Instead of Eq.~(\ref{limits-form}), one may assume the following form:
\be
g(k_{D}^{2}) \sim h(k_{D}^{2})=1-e^{-k_{D}^{2}/M^{2}}\,.
\ee

Then the same graviton propagator in Eq.~(\ref{eq:rara}) can be recast in the form
\ba \label{eq:loukoumi}
&& \Pi(k^{\mu})  \sim  \LF \mathcal{P}^{(2)}-\frac{1}{(D-2)A(k_{D}^{2})}\mathcal{P}_{s}^{(0)} \RF \non
&& \times \frac{1}{k_{N}^{2}+h(k_{D}^{2})(k_{D}^{2}-k_{N}^{2})}\,, ~~~~~
\ea
where we see that, in the IR, $A(k_{D}^{2}) \ra \frac{N-2}{D-2}$. Therefore in the IR limit the graviton propagator 
probes the reduced $N$ space-time dimensions. On the other hand, in the deep UV, $A(k_{D}^{2}) \ra 1$, the graviton 
propagator probes the original $D$ dimensions. Similar steps are followed for the denominator of Eq.~(\ref{eq:loukoumi}).

Now, there is a huge contrast between traditional KK-reduction and the prescription we have provided here. In the KK 
reduction, from higher dimensions in UV to lower dimensions in IR, there appear extra states which are known as the KK states. Furthermore, the dimensional reduction does not lead to the Einstein-gravity in the lower dimensions, the overall volume of the
extra dimensions appear as a scalar mode, known as a dilaton. This gives rise to a scalar-tensor theory, which becomes 
equivalent to an $F(R)$ gravitational theory at a classical level, rather than just the Einstein-Hilbert action~\footnote{In principle, one 
can recast the propagator of a graviton sector including the KK modes ($D \ra N$ dimensions from UV to IR) as
$\Pi_{KK}  \sim 
\Pi_{EH}+\sum_{\vec{n}}\frac{2\mathcal{P}^{(2)}}{k^{2}+m_{\vec{n}}^{2}}$,
where 
we have used the transversal projectors, see Eq.~\eqref{theta}, in the definition of $\mathcal{P}^{(2)}$ on shell, {\it i.e.}, 
$\theta_{\mu \nu}=\eta_{\mu \nu}-\frac{k_{\mu}k_{\nu}}{k^2} \ra \eta_{\mu \nu} + \frac{k_{\mu}k_{\nu}}{m_{\vec{n}}^{2}} \equiv P_{\mu \nu}^{\vec{n}}$.}.

The crucial difference between our approach and KK theories is that the KK modes indicate physical states in lower dimensions, or in IR, which correspond to massive particles, while in our approach, we obtain complex poles, whose origin will be  provided below.

\section{Interpretation of complex poles}\label{sec:rukruk}

Let us now analyse our scenario by asking how the graviton can be projected to probe certain degrees of freedom in a 
fixed $D$ dimensional space-time. This can be understood by analysing the poles in the graviton propagator appearing 
at the {\it intermediate} energies of $k^2_{D}$ in the Minkowski space rather than Euclidean space.

\begin{itemize}
\item{From $D\ra N$ (UV $\ra$ IR ):

Suppose we wish to find the poles of a graviton propagator which is $D$-dimensional in the UV and $N$-dimensional in the IR. This is the case very similar to KK-reduction, we
will be reverting to Minkowski space so that we can investigate the nature of the poles of the propagator. Then, by inspecting the denominator of Eq.~(\ref{eq:loukoumi}), we have
\be \label{eq:poles}
k_{N}^{2}+(1-e^{-k_{D}^{2}/M^{2}})(k_{D}^{2}-k_{N}^{2})=0\,,
\ee 
which yields
\be \label{eq:lolo}
\qquad~~~~ k_{0}^{2}=k_{1}^{2}+\dots+k_{D-1}^{2}-M^{2} W\left(\frac{k_{D}^{2}-k_{N}^{2}}{M^2}\right)\,,
\ee
where $W(z)$ is the Lambert $W$-function and $w=W(z)$ is the solution to $we^{w}=z$.  The Lambert $W$-function is a 
multivalued function~\cite{functions}, as $\log (z)$ is. Since the spatial components of $k_{D}^{\mu}$ and $k_{N}^{\mu}$ have to be real,  $k_{D}^{2}-k_{N}^{2}$ is a non-negative real number. Hence, the argument of the Lambert $W$-function in Eq.~\eqref{eq:lolo} is also a non-negative real number, 
meaning that we get infinitely many solutions, most of which are complex. These complex solutions have negative real part, meaning that we can write the complex solutions in the form $k_{D}^{2}=R+iS$, where $R$ is negative. Hence, for the complex solutions, $\mathrm{Re}(k_{D}^{2})=R<0$, indicating that the complex solutions are non-tachyonic from a higher-dimensional point of view (UV point of view); if they were to be tachyonic, we should have $R>0$.}

 \item{From $N\ra D$ (UV $\ra$ IR ):

Now, suppose we have the converse situation, and we wish to find the poles for the graviton propagator, again we are working in Minkowski space, which is $N$-dimensional in the UV and $D$-dimensional in the IR. Then by inspecting the denominator of Eq.~(\ref{eq:rara}), we obtain:
\be \label{eq:poles2}
k_{N}^{2}+e^{-k_{D}^{2}/M^{2}}(k_{D}^{2}-k_{N}^{2})=0\,,
\ee
which gives rise to
\ba \label{eq:qoqo}
&&k_{0}^{2}=k_{1}^{2}+\dots+k_{N-1}^{2}\non
&&-M^{2} W\left(-\frac{(k_{D}^{2}-k_{N}^{2})e^{-(k_{D}^{2}-k_{N}^{2})/M^{2}}}{M^2}\right)\,;~~~~~~
\ea
again we have infinitely many solutions, most of which are complex, since the argument of the Lambert $W$-function in Eq.~\eqref{eq:qoqo} is a non-positive real number (which follows from the fact that $k_{D}^{2}-k_{N}^{2}$ is a non-negative real number). Again, these complex solutions have negative real part, meaning that we can write these complex solutions in the form $k_{N}^{2}=R+iS$, where $R$ is negative. Hence, for these complex solutions, $\mathrm{Re}(k_{N}^{2})=R<0$, indicating that the complex solutions are non-tachyonic from a lower-dimensional point of view (UV point of view). Analysing the complex poles from a $D$-dimensional point of view is technically hard in this case.}

\end{itemize}

Interestingly, these complex poles appear in this prescription only at intermediate energies. The theory in the deep UV (in Euclidean space, $k^2_{D}\ra \infty$) and in the far IR (in Euclidean space, $k^2_{D}\ra 0$) has well-defined real massless graviton states, probing either $D$ or $N$ dimensions depending on the scenarios we are discussing.  One speculative interpretation would be that, at intermediate energies, the infinitely many complex poles would correspond to non-tachyonic unstable massive particles that decay to the two limiting cases, {\it i.e.}, to the massless graviton states in the IR and the UV. Indeed, this is an intriguing feature of projecting the graviton 
on certain spatial dimensions.  However, a clear physical interpretation in this context requires further investigation.

\section{Discussion and Conclusion}

The consequences of our analysis is indeed very interesting. Starting from $D$ dimensions in the UV, we could make the graviton propagator behave as $N$-dimensional in the IR and vice versa. As an example, in the UV for $k_{4}^{2}\ra \infty$, we could obtain the graviton propagator to mimic $3$- or  $2$-dimensional gravity, alike. In either cases, the theory of gravity indeed simplifies 
and perhaps avoids some of the thorny issues of gravity, which plague it in $4$ dimensions. Nature, fundamentally could survive in $2$ or $3$ dimensions in the UV, while in the IR bringing a new gravitational scale, $4$-dimensional $M_p$. 

Before we conclude, we wish to discuss some new avenues; so far, we have shown that the graviton propagating structure could be deformed continuously from $4 \ra 2$ dimensions by {\it entire functions} $g(k_4^2)$ or $h(k_4^2)$ (or, more generally, from $D \ra N$ dimensions). Of course, this analysis does not shed light on the precise form, in full generality, of these {\it entire functions}. Hopefully, there could be a novel way to determine the form of these functions by studying the cosmological implications - which would lead to an emergence of gravity from lower dimensions to higher dimensions, as postulated by many, see the review~\cite{Ambjorn}, and references therein. 

To summarise, we have shown that it is indeed possible to deform continuously the graviton propagator in the UV such that the graviton effectively sees the lower dimensions, by modifying the spin-$2$ and spin-$0$ projection operators deep within the UV, {\it i.e.}, 
$k_{D}^2\ra \infty$. The ramifications of this continuous deformation of graviton propagator also leads to the appearance of complex poles. One could speculate that these complex poles correspond to unstable decaying massive states; however, this conjecture requires further investigation.

\section{Acknowledgements}
AM would like to thank Robin Tucker,  Alessandro Codello and Giulio D' Odorico for discussions.
AM is supported by the STFC grant ST/J000418/1 and ST is supported by a scholarship from the Onassis Foundation.

\end{document}